\documentclass[draft]{agujournal2019}
\usepackage{url} %this package should fix any errors with URLs in refs.
\usepackage{lineno}
\usepackage[inline]{trackchanges} %for better track changes. finalnew option will compile document with changes incorporated.
\usepackage{soul}

\usepackage{amsmath} % added by me

\draftfalse

\journalname{Geophysical Research Letters}

\begin{document}

%%%%%%%%%%%%%%%%%%%%%%%%%%%%%%%%%%%%%%%%%%%%%%%
%  TITLE
%
% (A title should be specific, informative, and brief. Use
% abbreviations only if they are defined in the abstract. Titles that
% start with general keywords then specific terms are optimized in
% searches)
%
%%%%%%%%%%%%%%%%%%%%%%%%%%%%%%%%%%%%%%%%%%%%%%%

% Example: \title{This is a test title}

\title{Self-Affine Scaling of Earth's Islands}

%%%%%%%%%%%%%%%%%%%%%%%%%%%%%%%%%%%%%%%%%%%%%%%
%
%  AUTHORS AND AFFILIATIONS
%
%%%%%%%%%%%%%%%%%%%%%%%%%%%%%%%%%%%%%%%%%%%%%%%

% Authors are individuals who have significantly contributed to the
% research and preparation of the article. Group authors are allowed, if
% each author in the group is separately identified in an appendix.)

% List authors by first name or initial followed by last name and
% separated by commas. Use \affil{} to number affiliations, and
% \thanks{} for author notes.
% Additional author notes should be indicated with \thanks{} (for
% example, for current addresses).

% Example: \authors{A. B. Author\affil{1}\thanks{Current address, Antartica}, B. C. Author\affil{2,3}, and D. E.
% Author\affil{3,4}\thanks{Also funded by Monsanto.}}

\authors{Matthew Oline \affil{1}, Jeremy Hoskins \affil{2}, David Seekell \affil{3}, Mary Silber \affil{2}, B. B. Cael \affil{4}}

\affiliation{1}{Computational and Applied Mathematics, University of Chicago, Chicago, IL 60637, USA}
\affiliation{2}{Department of Statistics and Committee on Computational and Applied Mathematics, University of Chicago, Chicago, IL 60637, USA; NSF-Simons National Institute for Theory and Mathematics in Biology, Chicago, IL 60611, USA}
\affiliation{3}{Atle Fund Management, Stockholm, Sweden}
\affiliation{4}{Department of the Geophysical Sciences, University of Chicago, Chicago, IL 60637, USA; Climate Systems Engineering Initiative, University of Chicago, Chicago, IL 60637, USA; Institute of Climate and Sustainable Growth, University of Chicago, Chicago, IL 60637, USA}

%\affiliation{=number=}{=Affiliation Address=}
%(repeat as many times as is necessary)

% Corresponding author mailing address and e-mail address:

% (include name and email addresses of the corresponding author.  More
% than one corresponding author is allowed in this LaTeX file and for
% publication; but only one corresponding author is allowed in our
% editorial system.)

% Example: \correspondingauthor{First and Last Name}{email@address.edu}

\correspondingauthor{Matthew Oline}{moline@uchicago.edu}

%%%%%%%%%%%%%%%%%%%%%%%%%%%%%%%%%%%%%%%%%%%%%%%
% KEY POINTS
%%%%%%%%%%%%%%%%%%%%%%%%%%%%%%%%%%%%%%%%%%%%%%%
%  List up to three key points (at least one is required)
%  Key Points summarize the main points and conclusions of the article
%  Each must be 140 characters or fewer with no special characters or punctuation and must be complete sentences

% Example:
% \begin{keypoints}
% \item	List up to three key points (at least one is required)
% \item	Key Points summarize the main points and conclusions of the article
% \item	Each must be 140 characters or fewer with no special characters or punctuation and must be complete sentences
% \end{keypoints}

\begin{keypoints}
\item We construct a dataset of elevation profiles of over 130,000 islands whose areas span approximately 8 orders of magnitude.
\item Different geometric properties (volume, perimeter, and maximum height) scale with area according to different estimated fractal dimensions.
\item Estimates of fractal dimension are ordered according to the expected influence of erosion at the shoreline.
\end{keypoints}

%%%%%%%%%%%%%%%%%%%%%%%%%%%%%%%%%%%%%%%%%%%%%%%
%
%  ABSTRACT and PLAIN LANGUAGE SUMMARY
%
% A good Abstract will begin with a short description of the problem
% being addressed, briefly describe the new data or analyses, then
% briefly states the main conclusion(s) and how they are supported and
% uncertainties.

% The Plain Language Summary should be written for a broad audience,
% including journalists and the science-interested public, that will not have 
% a background in your field.
%
% A Plain Language Summary is required in GRL, JGR: Planets, JGR: Biogeosciences,
% JGR: Oceans, G-Cubed, Reviews of Geophysics, and JAMES.
% see http://sharingscience.agu.org/creating-plain-language-summary/)
%
%%%%%%%%%%%%%%%%%%%%%%%%%%%%%%%%%%%%%%%%%%%%%%%

%% \begin{abstract} starts the second page

\begin{abstract}
Earth’s relief is approximately self-affine, meaning a zoom-in on a small region looks statistically similar to a large region upon rescaling. Fractional Brownian surfaces give an idealized self-affine model of Earth’s relief with one parameter, the Hurst exponent $H$, characterizing the roughness of the surface. We compile a large dataset of topographic profiles of islands (N=131,063 with the range of areas covering approximately 8 orders of magnitude) and obtain four estimates for the Hurst exponent of Earth’s surface by fitting four statistical laws from the theory of self-affine surfaces concerning islands: (i) distribution of areas, (ii) volume-area relationship, (iii) perimeter-area relationship, and (iv) maximum height-area relationship. The estimated Hurst exponents indicate different fractal scaling behavior for different geometric features, and are sorted in order of increasing expected influence of coastal processes. This sheds light on the impact of coastal erosion and sedimentation on island geomorphology.

\end{abstract}

\section*{Plain Language Summary}
Several geological features have been observed to behave similarly to fractals; the classical example is the coastline paradox. A fractional Brownian surface is a simple mathematical model for a random surface that possesses a fractal-like property called self-affinity. In this model, islands are represented by regions of positive height in the random surface, with the regions of negative height thought of as being filled in with water. We consider four statistical laws arising from the theory of fractional Brownian surfaces which describe the distribution of the areas of these ‘islands’ and their relationship with other measured quantities of interest: the volume, the perimeter, and the maximum height. To see how closely islands on Earth follow these statistical trends, we take geometric measurements of a large number of islands across the globe using satellite elevation data, and compare. For a true self-affine surface, the four statistical laws can be expressed in terms of one parameter describing how rough the surface is. For Earth’s islands, the estimated roughness parameter is different when looking at different geometric features of islands, potentially informing our understanding of how natural processes like coastal erosion affect different geological features in different ways.

%%%%%%%%%%%%%%%%%%%%%%%%%%%%%%%%%%%%%%%%%%%%%%%
%
%  BODY TEXT
%
%%%%%%%%%%%%%%%%%%%%%%%%%%%%%%%%%%%%%%%%%%%%%%%

%%% Suggested section heads:
% \section{Introduction}
%
% The main text should start with an introduction. Except for short
% manuscripts (such as comments and replies), the text should be divided
% into sections, each with its own heading.

% Headings should be sentence fragments and do not begin with a
% lowercase letter or number. Examples of good headings are:

% \section{Materials and Methods}
% Here is text on Materials and Methods.
%
% \subsection{A descriptive heading about methods}
% More about Methods.
%
% \section{Data} (Or section title might be a descriptive heading about data)
%
% \section{Results} (Or section title might be a descriptive heading about the
% results)
%
% \section{Conclusions}

\section{Introduction}
%Text here ===>>>

Earth’s landmasses are shaped by complicated geological processes such as plate tectonics, volcanic activity, erosion, and sedimentation \cite{fossen2024,salles2023}. Examining the shapes and sizes of Earth’s landmasses can offer insight into these underlying processes and the evolution of Earth’s surface over time. Additionally, biodiversity and other ecosystem properties of islands are impacted by their size \cite{wardle1997,lomolino2000}. \citeA{korcak1940} observed that the sizes of certain geological features (among them the areas of islands, the areas of lakes, and the lengths of rivers) were right-skewed, with smaller features more abundant than larger ones. \citeA{frechet1941} found the power-law (or Pareto) distribution to be a good fit to such data. Power-law size distributions are frequently found in statistical physics models that exhibit self-similarity or invariance of scale, such as models of percolation, phase transitions, or self-organized criticality \cite{isichenko1992,bak1987}. Another hint at the fractal nature of Earth’s surface is the well-known coastline paradox \cite{mandelbrot1967,mandelbrot1983}: winding coastlines have features at all length scales, so the length of a coastline is not well-defined without also specifying a measurement length scale. \citeA{mandelbrot1975} proposed a fractional Brownian surface as a phenomenological fractal-like model for natural landscapes. Simulated landscapes are visually convincing (though this of course is subjective), and have found use in computer graphics \cite{fournier1982,barnsley1988}. Here we are particularly interested in islands; in the setting of fractional Brownian surfaces, one can imagine filling in the regions of negative height with water so that zero represents sea level and the regions of positive height represent islands sticking out from the ocean. From this idealized null model one obtains predictions about how certain measured quantities of islands scale with one another, which are described in section \ref{sec:theory}.

Efforts have been made to quantitatively assess the degree to which statistical scaling laws from the theory of self-affine surfaces match topographical data from Earth’s surface. One can obtain a local estimate of the Hurst exponent (or equivalently the fractal dimension) by fitting a variogram to elevation data \cite{mark1984,mcclean2000}, which describes the variation in elevation difference at two points as a function of the distance between the points. A Hurst exponent for the entirety of Earth’s surface has been estimated by fitting the cumulative distribution function of the areas of Earth’s islands, which is one of the four statistical laws we examine in this work. Kor\v{c}\'ak’s empirical number-area rule states that the number of islands with area greater than $A$ is proportional to $A^{-K}$, where $K$ was estimated to be $\sim0.65$ \cite{mandelbrot1975}. Mandelbrot observed that the ‘islands’ in a self-affine fractional Brownian surface follow an identical law, with $K = (2-H)/2$, which yields $H\sim0.7$ for the distribution of Earth's islands. The primary focus of this work is to fit three additional self-affine trends concerning the distribution of islands in order to understand how the estimated value of $H$ may vary depending on which geometric features are considered.

Our study is motivated in part by recent work on the geomorphological scaling behaviors of Earth's lakes, along with the curious symmetry between `lakes' and `islands' in theoretical self-affine surfaces. Just as islands can be represented as regions of positive height in a random surface, bodies of water can be represented as regions of negative height and are thus, in a sense, upside-down islands \cite{mandelbrot1983,russ1994}. This is only strictly true for theoretical random surfaces; on Earth, lakes and other bodies of water do not generally lie at sea level, are embedded in local depressions on a landmass, are often hydrologically connected, and are affected differently by geomorphological processes like sedimentation and erosion than islands are. It is not known how these material differences between Earth's lakes and islands produce differences in their scaling behaviors. By fitting the volume-area relationship for Earth's lakes, \citeA{caelvollakes} estimated a Hurst exponent of $H=0.4$, as predicted from some topographic estimates of Earth's $H$ \cite{mark1984,mcclean2000}. A near-identical $H$ was estimated by fitting the maximum height-area relationship \cite{caelmaxlakes}, indicating robustness of the estimate. However, \citeA{caelsizelakes} also showed that the distribution of areas of lakes and their perimeter-area relationship instead followed scaling laws predicted by percolation theory, exemplifying how interactions of natural processes can produce multifaceted scaling behaviors. In the case of lakes, horizontal scaling relationships correspond to random placement of lake water and vertical scaling relationships correspond to the topographic variations on which this lake water is embedded. Might islands show a similar complexity, whereby self-affine topographic variations and geomorphological processes interact to produce distinct scaling relationships for different geometric features?

\section{The Dataset}

\begin{figure}[ht]
\centering
\includegraphics[width=\textwidth]{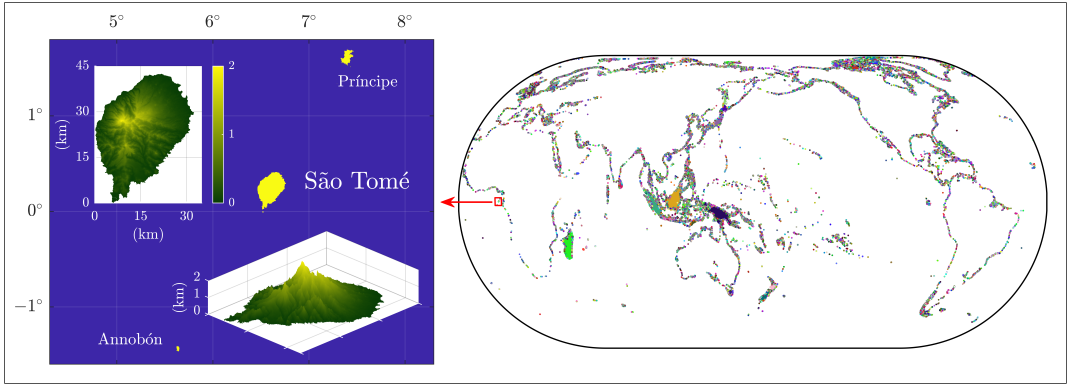}
\caption{\textit{(right)} Eckert IV (equal-area) projection of all islands in our constructed dataset. Pixels considered part of an island are plotted with uniformly sized circles to make small islands easily visible, circles are randomly colored according to connected component (island). Red square indicates location of the zoom-in in the left panel. \textit{(left)} Mercator projection of $[1.6^\circ S, 1.8^\circ N]$$\times$$[4.3^\circ E, 8.3^\circ E]$, with yellow indicating islands and blue representing ocean. Inlays show top-down view and 3-dimensional surface plot of S\~{a}o Tom\'{e} Island, with axis units in kilometers. In the surface plot, the height is exaggerated by a factor of 5 to show detail. }
\label{fig:map}
\end{figure}

 Landmasses were identified as connected components in the ASTER Global Digital Elevation Map V003 \cite{astergdem}, which has a 1-arcsecond pixel resolution ($\sim$30m near the equator) and spans 83 S to 83 N. Height values are integers in meters relative to the WGS84/EGM96 geoid. The components were filtered by size as well as location. Islands smaller than 0.01 km$^2$ were discarded due to poor resolution and very large landmasses were discarded due to practical memory constraints; the only large landmasses excluded were the continents, Greenland, and Baffin Island. We note that while the dataset includes islands larger than Baffin Island by area (specifically New Guinea, Borneo, and Madagascar), the Mercator projection of Baffin Island is larger due to area distortion at the high latitudes where it is located, thus it could not be included. Islands south of 60 S were discarded due to persistent cloud cover \cite{hall2006modis}. The area, volume, maximum height, and discretized perimeter were computed for the selected islands; see Supporting Information for full details on data processing. The final assembled dataset consists of N=131,063 discretized elevation profiles of islands from across the globe. The areas of islands in the dataset range from $0.01$ km$^2$ to $7.75\times 10^{5}$ km$^2$; the largest island is New Guinea, which is also the tallest in the dataset at 4833 m. The median area is 0.07 km$^2$, and the combined area is $7.27\times 10^{6}$ km$^2$, covering approximately 1\% of the Earth's surface. The dataset also contains the area and discretized perimeter of each island after `contracting' (removing the pixels on the boundary) or `expanding' (adding a layer of pixels to the boundary). These quantities give an estimate of the uncertainty due to discretization error, and are used in the regressions in Section \ref{sec:results}. We observe from Figure \ref{fig:map} that a large number of islands are located very close to a continent; this phenomenon is present in self-affine objects. Intuitively, a fractional Brownian motion contains noise at all frequencies. Near a zero-crossing, the presence of high frequency noise results in many nearby dips above and below zero, leading to many small 'coastal' islands.

\section{Empirical Analysis}
\label{sec:analysis}

\subsection{Theory}
\label{sec:theory}

Fractional Brownian surfaces are a family of random surfaces indexed by one parameter, the Hurst exponent $H \in (0,1)$, which describes how rough the surfaces are. Smaller $H$ corresponds to rougher surfaces and larger $H$ corresponds to smoother surfaces. $H=1/2$ corresponds to a standard Brownian surface; in one dimension this is a standard Brownian motion. These surfaces are self-affine, which is a slight relaxation of self-similarity: a self-affine surface is statistically identical to itself under a rescaling, where the rescaling may be different along different axes. Specifically, a two-dimensional fractional Brownian surface $b$ with Hurst exponent $H$ satisfies 
\begin{linenomath*}
\begin{equation}
b(x,y) \stackrel{d}{=} a^{-H}b(ax,ay) 
\end{equation}
\end{linenomath*}
for any scaling factor $a>0$, where the equality is in distribution. $H$ is algebraically related to the fractal dimension D; it can be shown that the surface $b(x,y)$ has fractal dimension $3-H$, and each level set $b(x,y) = c$ has fractal dimension equal to $2-H$. 

We consider four statistical laws concerning islands (regions of positive height) in self-affine surfaces: (i) the distribution of the areas, (ii) the volume-area relationship, (iii) the discretized perimeter-area relationship, and (iv) the distribution of the maximum height normalized by area. The laws are stated below and a brief explanation of each is given. For more details, see \citeA{isichenko1991,matsushita1991}. In what follows, $a$ represents the area of an island, $v$ the volume, $p$ the discretized perimeter, and $m$ the maximum height. 

\begin{linenomath*}
\begin{equation*}
\mathrm{(i)\quad} Pr(a>A)\sim A^{-k_1}, \quad k_1 = (2-H)/2
\end{equation*}
\end{linenomath*}

The areas of islands are power-law distributed: the probability that an island has area exceeding $A$ is proportional to $A^{-D/2}$, where $D=2-H$ is the fractal dimension of the level sets of the random surface.

\begin{linenomath*}
\begin{equation*}
\mathrm{(ii)\quad} v \sim a^{k_2}, \quad k_2 = (2+H)/2
\end{equation*}
\end{linenomath*}

Rescaling an island by a characteristic length scale $l$ should cause its area $a$ to be scaled by $l^2$, while its mean height $\bar{z}$ is scaled by $l^H$ due to the self-affine property. Thus the volume $v=a\bar{z}$ should scale like $l^{2+H}$, so $v\sim a^{(2+H)/2}$.

\begin{linenomath*}
\begin{equation*}
\mathrm{(iii)\quad} p \sim a^{k_3}, \quad k_3 = (2-H)/2
\end{equation*}
\end{linenomath*}

The set of disconnected coastlines of islands is a level set of a self-affine surface and has fractal dimension $D=2-H$. This means that for a fixed measurement resolution, rescaling by $l$ should cause an observed increase in measured perimeter $p$ by a factor of $l^D=l^{2-H}$. Since $a\sim l^2$, it follows that $p\sim a^{(2-H)/2}$.

\begin{linenomath*}
\begin{equation*}
\mathrm{(iv)\quad} y:=m/(\beta a^{H/2}) \mathrm{\ distributed\ according\ to\ density\ } f_H(y), \ \beta \mathrm{\ a\ free\ parameter}
\end{equation*}
\end{linenomath*}

Using a perturbative approach from a standard Brownian motion, \citeA{delorme2016} derived an expression for the probability density function of the maximum height of a one-dimensional fractional Brownian motion on an interval of length $l$. At leading order in $\varepsilon = H-1/2$, the normalized maximum height $y = m/l^H$ is distributed according to 
\begin{linenomath*}
\begin{equation}
f_H(y) \propto \sqrt{\frac{2}{\pi}}e^{-y/2}y^{1/H-2}e^{(H-1/2)(4ln(y)+\mathcal{G}(y))},\quad y>0
\end{equation}
\end{linenomath*}
 where
\begin{linenomath*}
\begin{align*}
\mathcal{G}(y) = \frac{y^4}{6} \,&_2F_2\left(1,1;\frac{5}{2},3;\frac{y^2}{2}\right)-3y^2+\pi(1-y^2)\textrm{erfi}\left(\frac{y}{\sqrt{2}}\right)\\
&+\sqrt{2\pi}e^{y^2/2}y+(y^2-2)(\gamma_E+ln(2y^2))
\end{align*}
\end{linenomath*}
and the constant of proportionality is such that $f_H$ integrates to one. $_2F_2$ is the generalized hypergeometric function, erfi is the imaginary error function, and $\gamma_E$ is the Euler-Mascheroni constant; see \cite{abramowitz1964} for details on special functions. \citeA{caelmaxlakes} considered the maximum depth $m$ of lakes normalized by their two-dimensional area $a$ raised to the power $H/2$, and found that the distribution of $m/a^{H/2}$ closely followed the one-dimensional formula up to a rescaling. This rescaling is meant to account for the constant of proportionality between the square root of the two-dimensional area of a lake and the one-dimensional characteristic length scale over which it achieves its maximum; we call this free parameter $\beta$.

\subsection{Results}
\label{sec:results}

\begin{figure}[ht]
\includegraphics[width=\textwidth]{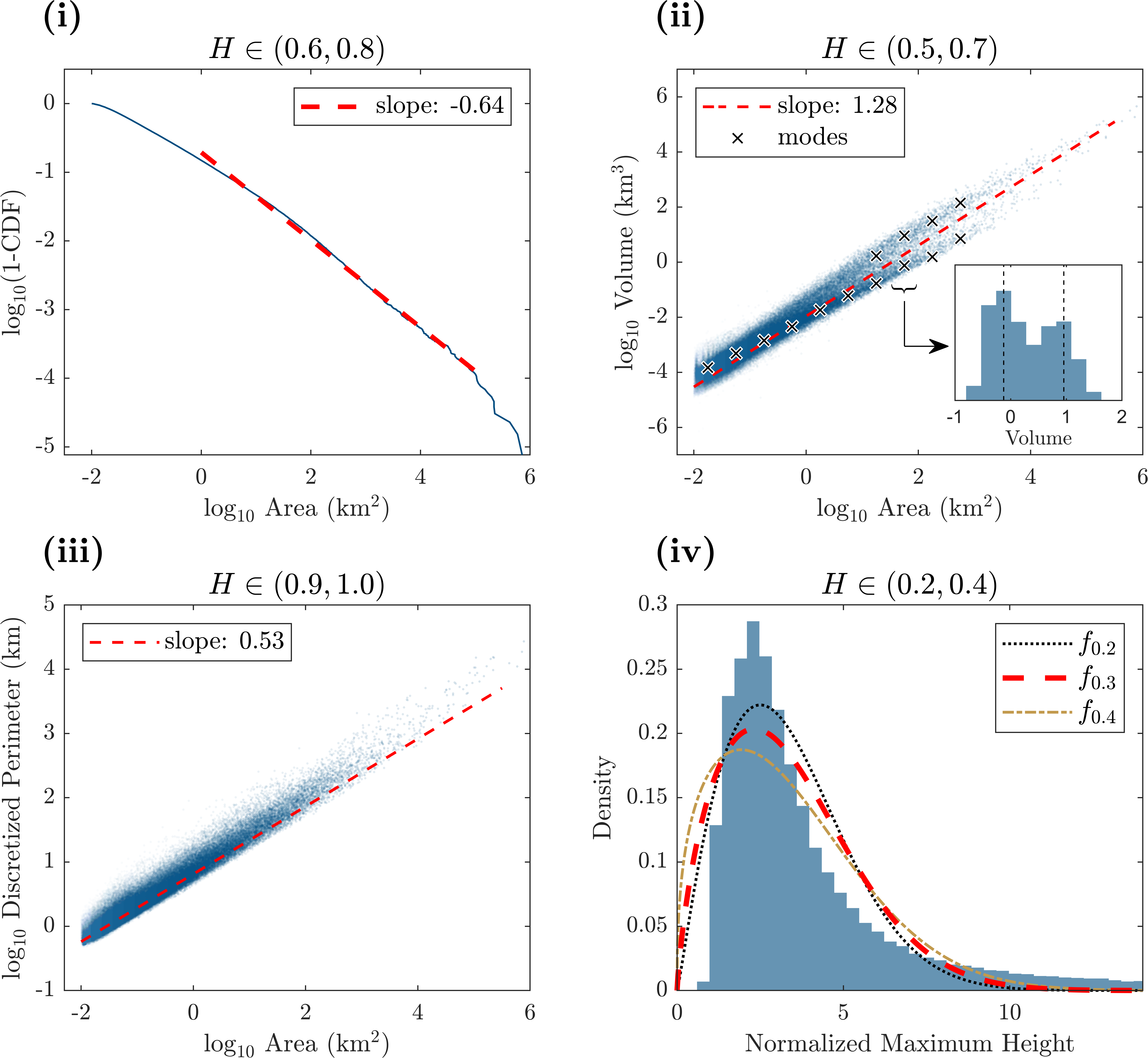}
\caption{\textit{(i)} Empirical complementary cumulative distribution function (equal to one minus the cumulative distribution function) of area in square kilometers.
\textit{(ii)} Scatterplot of volume in cubic kilometers vs area in square kilometers.
\textit{(iii)} Scatterplot of discretized perimeter in kilometers vs area in square kilometers.
\textit{(iv)} Histogram of normalized maximum (maximum height divided by area to the power $H/2$). Plots (i), (ii), and (iii), show the best fit line. Plot (iv) shows the best fit analytical probability density function $f_H$ for plot (iv), though we note in this case the fit is poor.
}
\label{fig:powerlaws}
\end{figure}

Between areas of $1$ and $10^5$ km$^2$, the islands appear to be very nearly power-law distributed according to (i), and we recover a similar slope to $k_1\sim 0.65$ reported by \citeA{mandelbrot1975}. This result is confirmed by varying the minimum area threshold to be considered in the fit then calculating the maximum likelihood estimator of the power-law exponent and the Kolmogorov-Smirnov error \cite{clauset2009}. The data is in good agreement with the power-law fit (with Kolmogorov-Smirnov statistic of around 0.04) for minimum areas between 1 and 100 km$^2$. Over this range of minimum areas, one obtains values of $k_1$ in the interval $(0.6,0.7)$, leading to estimates of $H$ in the interval $(0.6,0.8)$. However there is a slight but noticeable curvature in the empirical complementary cumulative distribution function, and the curvature becomes more pronounced for areas below $1$ km$^2$. This curvature over several orders of magnitude in length scales indicates a deviation from the expected power-law behavior. Above $10^5$ km$^2$ there is an insufficient number of large landmasses to get a visually smooth curve.

Relationships (ii) and (iii) were fit using a Type II York regression \cite{york2003} to account for measurement errors in both plotted variables. Estimates of the errors associated with discretization were obtained for each island by computing the area and perimeter of the island after adding a layer of pixels around the boundary and after removing a layer of pixels from the boundary. The approximately 8 orders of magnitude in area spanned by the islands in our dataset gives us a sufficiently large range of length scales over which to perform our analysis without explicitly taking into account finite size effects. This is further justified by the fact that the largest island in our dataset covers less than $0.5\%$ of Earth’s surface. The fit slopes were $1.28\pm 0.03$ and $0.52\pm 0.01$, giving estimates for $H$ of $0.57\pm0.06$ and $0.95\pm0.02$ respectively. Uncertainty was estimated using a $95\%$ confidence interval from bootstrap resampling. We make the interesting observation that for a fixed slice of sufficiently large areas, the marginal distribution of volumes is bimodal, indicating that large islands tend to fall into one of two distinct regimes: islands that have a noticeably larger volume than expected from their area, and islands that have a noticeably smaller volume than expected from their area. This bimodality shows no clear relationship with any geographic characteristics like latitude, longitude, ocean basin, or distance to continental coastline. Preliminary tests indicate that this clustering corresponds to known classifications of tall (high) and short (low) islands according to their geological makeup \cite{nunn2016}; see Discussion.

The density $f_H$ for the normalized maximum in (iv) was fit by minimizing the Kuiper statistic \cite{kuiper1960} (equal to the sum of the one-sided Kolmogorov-Smirnov statistics, i.e. the range of the difference of CDFs) compared to the data. The best fit tuple $(H,\beta)\approx(0.32,2.96)$ resulted in a Kuiper statistic of $\approx0.1$, suggesting that the data does not match the predicted trend. One can see that the densities around this range of $H$ plotted in panel (iv) of Figure \ref{fig:powerlaws} do not capture the tail behavior of the data. In fact, no $f_H$ decayed sufficiently slowly at the tail to closely match the data, even when attempting to fit only the data above the median. We conclude that in this case, there are more islands with very large maximum heights relative to their areas than what was predicted by the formula adapted from the one-dimensional case. Additionally, there is a lack of islands with very small maximum heights relative to their areas in the data, indicated by the mismatch of the data and the plotted densities near 0. Densities $f_H$ for $H\geq1/2$ fared even worse in this regard than those plotted, as they have positive density at 0.

\section{Discussion}
\label{sec:discuss}

Statistical analysis of our newly constructed dataset of topographic profiles of Earth’s islands demonstrated fractal-like scaling behaviors in the distribution of the areas of islands and the relationships between the area and other geometric features. The distribution of areas, the volume-area relationship, and the discretized perimeter-area relationship appeared to roughly conform to power-law predictions arising from the theory of self-affine surfaces, while the predicted relationship between the area and the maximum height of islands extrapolated from one-dimensional fractional Brownian motion was a poor fit. There is significant variation in the estimates of the Hurst exponent $H$ of Earth’s surface from the four relationships examined, even among the three cases where the data roughly obeyed the general form of the predicted relationship. Specifically, estimates of the Hurst exponent (or the fractal dimension) of Earth’s surface depend on the selection of geometric features of islands from which the estimates are obtained. This indicates that the one-parameter family of fractional Brownian surfaces indexed by $H$ is too simple a model to capture the complex scaling phenomena we observe in the data. However, the ways in which the data differs from the idealized self-affine model could help inform our understanding of the geomorphological processes that shape Earth's landscapes. Our four estimates for the Hurst exponent are sorted according to the expected influence of wave erosion on the selected geometric feature in relation to area. The smoothest estimate, $H \sim 0.95$, was obtained when considering the discretized perimeters, i.e. shorelines, which are subject to intense erosion from moving water. Next is $H \sim 0.7$ when considering the areas alone, followed by $H \sim 0.6$ when considering the volume. Intuitively, if erosion wears away a thin layer of mass from an island at the coastline, this would result in a larger relative decrease in the area than the volume due to the respective dimensionalities, so the smoothing effect would be greater for the areas than for the volumes, though both of these quantities would be less affected by wave erosion than the perimeter. Lastly, we expect the peaks of islands to be unaffected by erosion at the coastline (though wind erosion and mass wasting could affect the maximum height), so the decrease in area results in more spiky islands, and indeed we find the roughest estimate $H \sim 0.3$ from looking at the maximum height normalized by area. We reiterate, however, that while this last estimate is the best fit value of $H$ when considering the one-dimensional formula for the maximum height-area relationship, the formula is a poor fit to the data for all $H$. It is unclear at the moment how exactly these inconsistencies in $H$ across scaling laws may be related to scale-dependent deviations within individual scaling laws, such as the curvature observed in the complementary cumulative distribution function of island areas (Figure 2 (i)).

These observations offer a testable hypothesis to be explored in future work: can one replicate these scaling behaviors by first starting with islands generated from a self-affine surface and then simulating erosion of their shorelines? This idea of starting with a fractal landscape and simulating erosion has been used to produce more realistic simulated landscapes, complete with erosion features such as ridges and valleys \cite{musgrave1989}. Though our dataset is a snapshot of Earth's islands at the present time, the shape and sizes of islands are dynamic, slowly but constantly changing as a result of erosion and sedimentation. Individual islands of varying ages will have been subjected to these forces for different amounts of time and thus may have very different shapes, but the adherence to global self-affine scaling laws would be a result of global (approximate) scale-invariance of these shaping processes. We note, however, that self-affine scaling properties are not unique descriptors of the geomorphological processes shaping islands. For instance, ridge and valley networks carved by fluvial incision are ubiquitous in natural landscapes, though absent in self-affine random surfaces. Some continuum models of landscape evolution incorporating fluvial incision and hillslope transport mechanisms such as diffusive creep and mass wasting (also lacking in the self-affine surface model) give rise to fractal-like simulated landscapes \cite{bonetti2020,anand2023}. That fractal-like behavior can be obtained from models with vastly different underlying mechanisms illustrates the limitations in drawing conclusions from scaling behaviors alone.

\begin{figure}
\includegraphics[width=\textwidth]{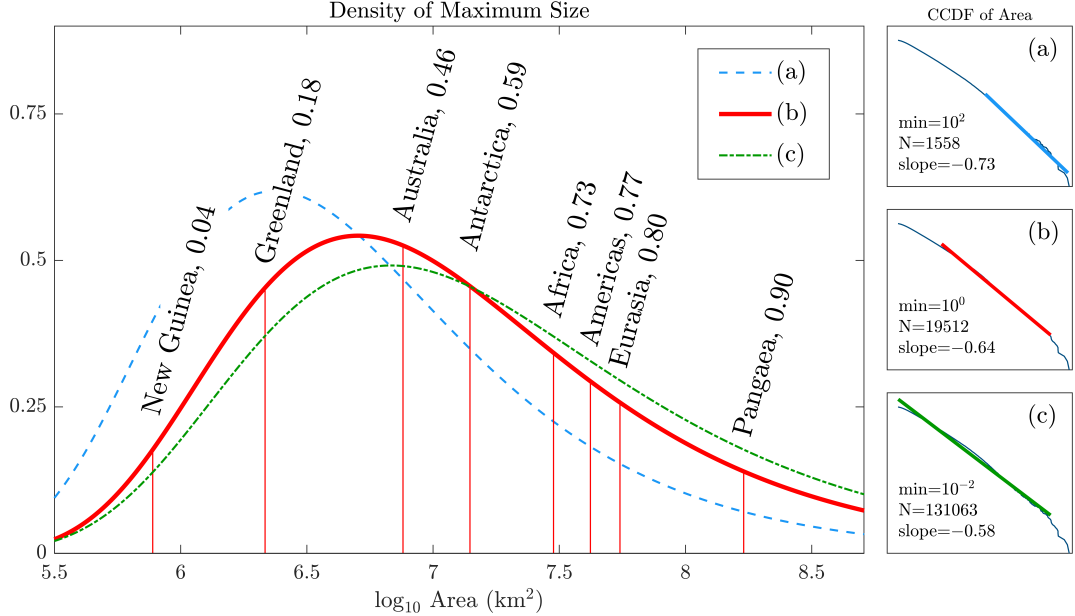}
\caption{(left) Probability density functions for the maximum area of a collection of $N$ landmasses assuming independently distributed areas from a true power-law distribution with the exponent fit from the empirical complementary cumulative distribution function of area. Sizes of notable landmasses (New Guinea (the largest island in our dataset), Greenland, the continents, and Pangaea) marked along with the value of the theoretical CDF at these points. (right) Three different fits of the CCDF for area with a range of fit slopes. $N$ is the number of landmasses in our dataset above the minimum area considered in each fit.}
\label{fig:cont_vs_isl}
\end{figure}

Though self-affine models of Earth's surface are phenomenological, simply capturing observed fractal-like properties without explaining their causes, identifying where self-affine scaling laws hold (or sometimes more interestingly, where they do not hold) may help answer questions about Earth’s landmasses, though one should keep in mind that there are obvious limitations on the range of length scales over which we can observe these self-affine scaling laws, given the finite size of the Earth and the limited resolution of available measurements. For instance, the distinction between islands and continents is somewhat blurred, with existing definitions using some combination of size, tectonic plate structure, and simple convention. But is there a discernible difference between islands and continents from the point of view of Earth’s self-affinity? We provide one answer to this question by a simple statistical test. Specifically, we consider the theoretical distribution of the maximum area over a collection of islands, assuming the island sizes are independently power-law distributed according to the fit we obtain from our dataset which does not include the continents, and ask how likely it would be that the maximum area is comparable to the areas of the continents. The results are displayed in Figure \ref{fig:cont_vs_isl}. We can see for example that 64\% of such collections of power-law distributed islands would have a maximum size at least as large as Australia. This suggests that Australia (and perhaps the other continents as well, depending on significance threshold) should not be considered an outlier in the distribution of areas of Earth's islands.

In the idealized self-affine model of Earth's surface, there is a duality between landmasses and bodies of water; one can think of connected regions of positive height as islands or connected regions of negative height as lakes, and they will have the same size distribution due to the $b(x,y) \leftrightarrow -b(x,y)$ symmetry of fractional Brownian surfaces. The results of this calculation regarding the sizes of continents as opposed to islands are in contrast to the case for Earth’s bodies of water, where the null hypothesis that oceans conform to the power-law distribution extrapolated from the areas of lakes was rejected \cite{caelpowerlaw}. Another interesting contrast is the disagreement between the maximum height-area relationship of islands and the formula adapted from one-dimensional fractional Brownian motion: in the case of Earth’s lakes, the maximum depth-area relationship closely matches the one-dimensional formula \cite{caelmaxlakes}. Additionally, the bimodality in the marginal distributions of volumes for sufficiently large areas (see the inset panel in Figure \ref{fig:powerlaws}(ii)) is a novel phenomenon not seen in lakes. Though certain aspects of both islands and lakes obey geometric scaling laws from the theory of self-affine surfaces, our analysis shows islands have markedly different behavior than upside-down lakes. The areas and locations of Earth's islands have been mapped extensively along with ecological properties such as as biodiversity, temperature, and precipitation \cite{sayre2019,weigelt2013}. 
However, less is known about the global distribution of island volumes. The dataset constructed here allows for new comparisons of the hypsometry of islands to the bathymetry of lakes in order to understand the different geological processes that shape Earth's surface above or below water. 

The dataset constructed here is ripe for further statistical analysis; of particular interest is the identified bimodality in the marginal distributions of volume at sufficiently high ranges of area.  The bimodality observed Figure \ref{fig:powerlaws}(ii) naturally divides the islands with large areas into two classes depending on their volumes. This is in agreement with known classifications of high/low islands based on height, which is strongly related to geological makeup. For example, the volcanic `high' Hawaiian Islands fall in the upper branch, and the limestone `low' Bahamas Islands fall in the low-volume branch. It remains to be seen if similar statistical clustering based on geometric features of islands corresponds to shared geological properties. However, detailed qualitative geological characterizations of islands likely require more case-by-case attention (especially where there is some ambiguity in the geological distinction), as opposed to purely geometric quantities which can be easily computed for a large number of islands. Further work is also needed for a comprehensive understanding of the relationship between the maximum heights of islands and their areas. For this and other comparisons to fractional Brownian surfaces, it may be instructive to employ a brute force computational approach. By generating a large number of fractional Brownian surfaces, one can obtain empirical distributions of quantities of interest as a function of the Hurst exponent, and compare to data.

\section*{Open Research Section}
\label{sec:openresearch}

The data used in this analysis is available on Zenodo \cite{oline2026}. See Supporting Information for a detailed description of the dataset.

\section*{Conflict of Interest Statement}

The authors have no conflicts of interests to disclose.

%This section MUST contain a statement that describes where the data supporting the conclusions can be obtained. Data cannot be listed as ''Available from authors'' or stored solely in supporting information. Citations to archived data should be included in your reference list. Wiley will publish it as a separate section on the paper’s page. Examples and complete information are here:
%https://www.agu.org/Publish with AGU/Publish/Author Resources/Data for Authors

\acknowledgments
J.G.H. was supported in part by a Sloan Research Fellowship. This research was supported in part by grants from the NSF (DMS-2235451) and Simons Foundation (MPS-NITMB-00005320) to the NSF-Simons National Institute for Theory and Mathematics in Biology (NITMB). 

%%%%%%%%%%%%%%%%%%%%%%%%%%%%%%%%%%%%%%%%%%%%%%%
% REFERENCES and BIBLIOGRAPHY
%
%\bibliography{<name of your .bib file>} don't specify the file extension
% don't specify bibliographystyle
%
%%%%%%%%%%%%%%%%%%%%%%%%%%%%%%%%%%%%%%%%%%%%%%%

%\bibliography{ enter your bibtex bibliography filename here }

\bibliography{refs}

%Reference citation instructions and examples:
%
% Please use ONLY \cite and \citeA for reference citations.
% \cite for parenthetical references
% ...as shown in recent studies (Simpson et al., 2019)
% \citeA for in-text citations
% ...Simpson et al. (2019) have shown...
%
%
%...as shown by \citeA{jskilby}.
%...as shown by \citeA{lewin76}, \citeA{carson86}, \citeA{bartoldy02}, and \citeA{rinaldi03}.
%...has been shown \cite{jskilbye}.
%...has been shown \cite{lewin76,carson86,bartoldy02,rinaldi03}.
%... \cite <i.e.>[]{lewin76,carson86,bartoldy02,rinaldi03}.
%...has been shown by \cite <e.g.,>[and others]{lewin76}.
%
% apacite uses < > for prenotes and [ ] for postnotes
% DO NOT use other cite commands (e.g., \citet, \citep, \citeyear, \nocite, \citealp, etc.).
%

\end{document}